\documentclass[12pt]{iopart}
\usepackage{color} \usepackage{newlfont}  \usepackage{epsfig}

\begin{document}
\title{Sub-micrometer epitaxial Josephson junctions for quantum circuits}
\author{Jeffrey S. Kline$^1$, Michael R. Vissers$^1$, Fabio C.
S. da Silva$^1$, David S. Wisbey$^1$\footnote{Present address: Department of Physics, Saint Louis University, Saint Louis, MO 63103, USA}, Martin Weides$^1$, Terence J. Weir$^2$, Benjamin Turek$^2$\footnote{Present address: The Johns Hopkins University Applied Physics Laboratory, 11100 Johns Hopkins Road, Laurel, MD 20723, USA}, Danielle A. Braje$^2$, William D. Oliver$^2$,
Yoni Shalibo$^3$, Nadav Katz$^3$, Blake R. Johnson$^4$, Thomas A. Ohki$^4$, and David P. Pappas$^1$}
\address{$^1$National Institute of Standards and Technology, Boulder, CO 80305, USA}
\address{$^2$MIT Lincoln Laboratory, 244 Wood Street, Lexington, MA 02420, USA}
\address{$^3$Racah Institute of Physics, The Hebrew University of Jerusalem, Jerusalem 91904, Israel}
\address{$^4$Raytheon BBN Technologies, Cambridge, MA 02138, USA}
\eads{\mailto{jeffrey.kline@NIST.gov}, \mailto{david.pappas@NIST.gov}}

\begin{abstract} We present a fabrication scheme and testing results for epitaxial sub-micrometer Josephson junctions. The junctions are made using a high-temperature (1170 K) ``via process'' yielding junctions as small as 0.8 $\mu$m in diameter by use of optical lithography. Sapphire (Al$_2$O$_3$) tunnel-barriers are grown on an epitaxial Re/Ti multilayer base-electrode. We have fabricated devices with both Re and Al top electrodes. While room-temperature (295 K) resistance versus area data are favorable for both types of top electrodes, the low-temperature (50 mK) data show that junctions with the Al top electrode have a much higher subgap resistance. The microwave loss properties of the junctions have been measured by use of superconducting Josephson junction qubits. The results show that high subgap resistance correlates to improved qubit performance.

Contribution of U.S. government, not subject to copyright. \end{abstract} \maketitle

\section{Introduction}

Josephson junction superconducting devices are promising candidates for qubits in quantum information circuits \cite{Clarke:review}. The tunnel-barriers in these junctions are typically amorphous AlO$_x$ made by room-temperature ($T = 295$ K) oxidation of thin films of aluminum. Qubit spectroscopy reveals coupling to stochastically distributed two-level systems (TLS) in the tunnel-barrier \cite{Simmonds:TLS}.
These TLS are observed as avoided level crossings (i.e., splittings) in the qubit spectroscopy. For the amorphous AlO$_x$ tunnel-barrier, the density of TLS splittings is measured to be 0.5 ($\mu$m$^2$GHz)$^{-1}$ \cite{Martinis:loss, Kline}. While the physical origin of TLS is still under debate, it is clear that their interaction with the qubit is detrimental because they can absorb energy and decohere the qubit state. These TLS have a random distribution in frequency space and coupling strength. Unless some strategy for reducing the number of TLS is used, it is highly likely that TLS splittings will appear close to the desired operation frequency when circuits with multiple qubits are constructed.

To date, there have been two strategies to reduce the number of TLS in qubit junctions. The first is to reduce the junction area as much as possible. Sub-micrometer Josephson junctions made by use of electron-beam lithography and Al shadow-evaporation \cite{Dolan77} are highly successful in charge qubits, transmons, flux qubits, and low-impedance flux qubits \cite{Kim:chgqubit, SchreierPRB08, Mooij, Steffen:Zqubit}. However, the absence of metal cross-overs in electron beam-defined circuits limits the available circuit designs (e.g., no gradiometric flux coils). Step-edge junctions fabricated by use of optical lithography for phase qubits can be made as small as 1 $\mu$m$^2$ \cite{Steffen:overlap, Weides_Trilayer}. While cross-overs are part of step-edge technology, multiple qubit circuits will still suffer from the residual stochastic TLS splitting distribution due to the high density of TLS splittings.

The second strategy is to reduce TLS density by use of epitaxial materials. While this typically involves high-temperature processing \cite{Oh:junctions}, it has yielded improved performance. Oh {\it et al.} observed an  $\sim$80\% reduction in the density of TLS in a large-area (70 $\mu$m$^2$) phase qubit with a crystalline Al$_2$O$_3$ tunnel-barrier when compared to amorphous AlO$_x$ \cite{Oh:TLS}. This technology uses optical lithography, and cross-overs are easily made.

Our goal in this work is to combine these two strategies by developing a process to reduce the size of epitaxial junctions for high-coherence qubits. In addition, we evaluate the efficacy of replacing the Al top electrode with Re. This is motivated by the discussion in \cite{Oh:TLS}, where it was hypothesized that the residual TLS may originate at the Al$_2$O$_3$-Al interface.  To test this hypothesis, we studied Re top-electrode junctions and qubits, and compared them to devices with Al top-electrodes.

In order to reduce the junction size, we first tried a standard trilayer process \cite{Oh:junctions} but the photoresist pillar washed away in the developer rinse for junction sizes smaller than $\sim$2 $\mu$m. While a trilayer process for sub-micrometer junctions does exist \cite{Berggren}, it requires chemical-mechanical planarization and this is not available in our facility. Instead, we developed a high-temperature ``via process'' similar to a scheme used for masking GaN nanowire growth \cite{Bertness}. As discussed below, the epitaxial base-electrode is grown on a high-quality substrate, an insulator with a via to the base-electrode is defined, and the epitaxial tunnel-barrier is then grown in the via after heating and recrystallizing the surface. We have measured both the room-temperature ($T = 295$ K) and low-temperature ($T < 100$ mK) properties of these single-junction devices, and we have also fabricated qubit devices and measured their performance.

\section{Substrate preparation and base-electrode growth} All of our devices were fabricated on single-crystal Al$_2$O$_3$(0001)
sapphire wafers. The wafers are 76.2 mm in diameter and 0.43 mm thick. As received from the manufacturer, the surface of the substrate exhibits no lateral crystalline order when imaged by atomic force microscopy (AFM), i.e., it is amorphous. To improve the crystalline order, we heat the substrate in a tube furnace to 1370 K for 20 hours in a 14:1 nitrogen-to-oxygen gas mixture at atmospheric pressure \cite{Cuccureddu,Yoshimoto}. After the furnace treatment, we observe atomic step terraces and lateral order. In addition, we find a correlation between the miscut angle (angle between dicing saw cut and the (0001) crystal plane) and terrace size measured by AFM: even a small, 0.3$^{\circ}$ miscut limits the terraces to $\sim$100 nm wide, while a nominal 0.0$^{\circ}$ miscut yields $\sim$390 nm wide terraces. However, we found that the surface morphology of the Re base-electrode is independent of furnace treatment and miscut angle.

\begin{figure}[ht] \centerline{\includegraphics[width = 13.97 cm]{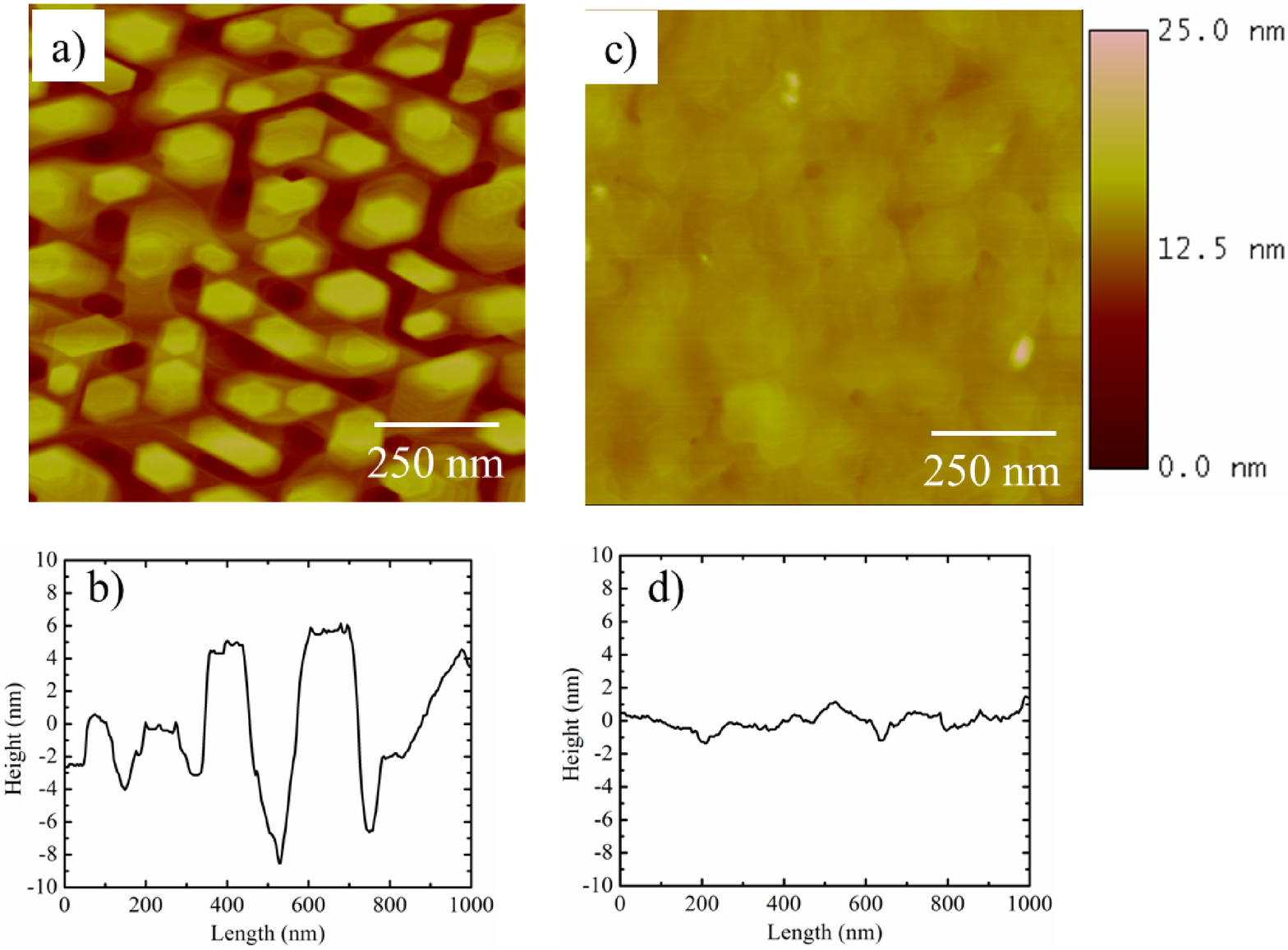}} \caption{Comparison of base-electrode film morphology. Rhenium base-electrode 165 nm thick, sputter-deposited at 1170 K: (a) AFM image (1$\times$1 $\mu$m$^2$) with rms roughness = 3.2 nm and (b) AFM linescan. (Re/Ti)$_{12}$Re multilayer base-electrode film 150 nm thick, sputter-deposited at 1170 K: (c) AFM image (1$\times$1 $\mu$m$^2$) with rms roughness = 0.6 nm and (d) AFM line scan. Z-scale is 25 nm for both images.} \label{rhenium}
\end{figure}

At first we used a 165 nm thick rhenium film for the base-electrode, deposited by use of ultra-high vacuum (UHV) DC sputtering \cite{Oh:rhenium}. The substrate temperature is held at 1170 K, the deposition rate is 3 nm/min, and the argon sputter
gas pressure is 0.7 Pa. For this and all subsequent layers, we rotate the substrate during deposition to improve film thickness
uniformity. The magnetron sputter guns are mounted in a sputter-up configuration at 35$^{\circ}$ off normal and 15 cm from the
substrate. Using this technique, we obtain crystalline rhenium films. We find that these film are characterized by $\sim$100 nm diameter hexagonal islands with $\sim$15 nm height, as shown in figure \ref{rhenium}(a). The root mean square (rms) roughness of the Re films is 3.2 nm, and is indicative of step-bunching and limited mobility of the Re during deposition. For comparison, a polycrystalline or crystalline Nb base-electrode suitable for high-quality Nb-Al/AlO$_x$-Nb Josephson junctions has roughness $\leq 0.5$ nm \cite{Du, Welander}.

In order to obtain a smoother Re surface for subsequent growth of the barrier, we found that it is possible to reduce the rms
roughness of the base-electrode film while maintaining crystallinity by using a Re/Ti multilayer. In this process, we
deposit a 10 nm Re layer and then cap it with 1.5 nm of Ti. Both films are UHV sputtered at 1170 K. Titanium has a lower
surface free energy (1.9 J/m$^2$) than rhenium (2.2 J/m$^2$) and acts as a wetting layer, resulting in a significantly
smoother surface, as shown in figures \ref{rhenium} (a) and (c). By repeating the Re/Ti unit cell structure twelve times and then
capping with a 10 nm Re top layer (i.e., (Re/Ti)$_{12}$Re), we obtain base-electrode films 150 nm thick with an rms roughness of only 0.6 nm. Line scans from the respective AFM images are shown in figures \ref{rhenium} (b) and (d), illustrating that the multilayer film is much smoother, with fewer vertical edges, than the pure Re film. Reflection high energy electron diffraction (RHEED) patterns (not shown) from pure Re and (Re/Ti)$_{12}$Re films are indistiguishable, indicating that base-electrode crystallinity is not degraded by using the Re/Ti multilayer. This base-electrode surface is much more favorable for tunnel-barrier growth with a sharp metal-insulator interface \cite{Du, Kohlstedt, Kominami}.

\section{Tunnel-barrier and top-electrode growth} 
\begin{figure}[ht] \centerline{\includegraphics[width = 13.97 cm]{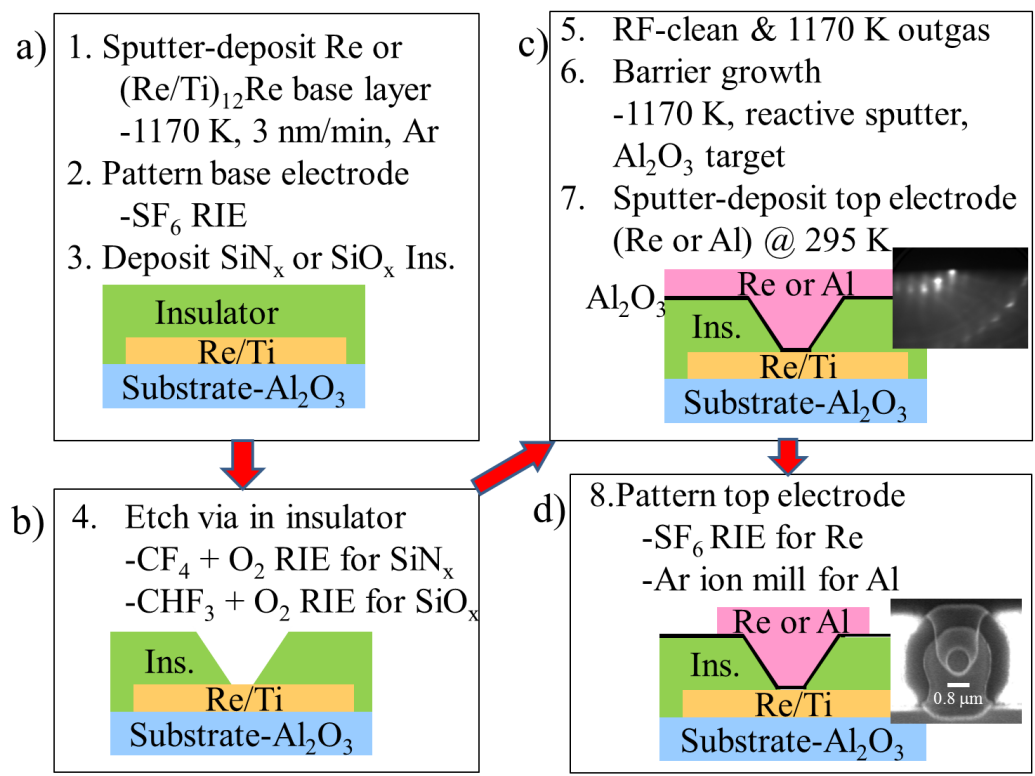}} \caption{Fabrication schematic of epitaxial Josephson junctions. (a) the (Re/Ti)$_{12}$Re base  electrode is patterned and encapsulated with insulator. (b) The via is etched into the insulator. (c) RF-clean, 1170 K outgas, tunnel-barrier growth, and top-electrode deposition are all performed in UHV without breaking vacuum. The inset shows a crystalline RHEED pattern of a Re base-electrode after 1170 K anneal. (d) The top-electrode is patterned. Inset shows an SEM image of a junction with 0.5 $\mu$m designed diameter (measured diameter = 0.8 $\mu$m). Additional wiring layers are needed for gradiometric devices. } \label{fabcartoon} \end{figure}

Once the base-electrode is grown, we proceed to define the tunnel junction and top-electrode by use of the via process. This process is illustrated in figure \ref{fabcartoon}. The tunnel-barrier and top-electrode are deposited after the insulator and via are defined. To accomplish this, we remove the wafer (with the epitaxial base-electrode already grown) from the UHV sputter tool. The first step of the process is shown in figure \ref{fabcartoon}(a), where the (Re/Ti)$_{12}$Re base-electrode is patterned by use of a 500 V SF$_6$ RIE at 2 Pa, etching all the way down to the substrate. The base-electrode is then covered with a 220 nm thick cross-over insulator, either SiO$_x$ or SiN$_x$, by use of plasma enhanced chemical vapor deposition at $T = 295$ K.

Vias are then etched in the insulator by use of a 240 V RIE, shown in figure \ref{fabcartoon}(b). We use CHF$_3$ + O$_2$ at 13 Pa for the SiO$_x$ insulator and CF$_4$ + O$_2$ at 2 Pa for SiN$_x$. This step defines the tunnel junction size and shape. This RIE has a 3:1 (5:1) selectivity in etch rate between SiO$_x$ (SiN$_x$) and Re, allowing us to stop fairly effectively when we reach the top of the crystalline Re base-electrode. Minimizing the over-etch into the base-electrode is critical, as it could create vertical walls around the edge of the via and it also could etch down to the Ti wetting layer. Tunnel-barrier coverage on these vertical sidewalls would be problematic and prone to pinholes and uneven coverage. We use a laser interferometer endpoint-detection scheme to minimize the over-etch (typically $< 10$ \% of the total etch time). We estimate the maximum amount of Re removed by the via over-etch and RF-clean (see below) is $\sim$5 nm. This leaves $\sim$5 nm of Re covering the nearest Ti wetting layer, so that our tunnel barriers are grown on Re, not Ti. This assertion is also confirmed by RHEED patterns: Re and Ti have distinct patterns.

After the via etch, the base-electrode in the bottom of the via has been amorphized due to the over-etch portion of the via etch. This surface needs to be cleaned and recrystallized before a tunnel junction can be grown on it. We do this by loading the wafer back into the
UHV sputter tool, performing an argon RF-clean to remove  $\sim$2 nm of material, and then heating the wafer to 1170 K for 1 hour as shown in figure \ref{fabcartoon}(c). RHEED image shown in the inset of \ref{fabcartoon}(c) and \emph{ex-situ} AFM images (not shown) indicate that the Re surface is clean and re-crystallized. The epitaxial Al$_2$O$_3$ tunnel-barrier is then grown {\it in situ} by use of UHV RF magnetron-sputtering from a sintered Al$_2$O$_3$ sputter target \cite{Barner1,Barner:tunneling}. For this deposition, the substrate temperature is held at 1170 K, the deposition rate is 0.9 nm/min, and the sputter gas pressure is 0.7 Pa argon with 5 mPa oxygen. The oxygen gas is necessary to prevent oxygen loss from the aluminum oxide at high-temperature and to obtain fully stoichiometric Al$_2$O$_3$. The thickness of the tunnel-barrier is monitored {\it in situ} by use of spectroscopic ellipsometry. We grow crystalline aluminum oxide films $1.8\pm0.2$ nm thick as tunnel-barriers. We find that they are conformal to the Re base-electrode, as evaluated by comparing AFM images and finding them to be indistinguishable from those of the base-electrode. In addition, we note here that other barrier growth conditions were explored, for example, growing at 1170 K without oxygen and also growth at $T = 295$ K followed by an 1170 K anneal in oxygen. These resulted in a low resistance$\times$area (RA)-product $<$ 400 $\Omega$$\mu$m$^2$ for tunnel-barriers up to 9 nm in thickness, and the RA-product was independent of barrier thickness. From this, we conclude that tunnelling was not the dominant transport mechanism for barriers grown without oxygen or at $T = 295$ K with an anneal. According to AFM and electrical-isolation measurements of metallic cross-overs, the SiO$_x$ and SiN$_x$ cross-over insulators are stable (i.e., no flowing of insulator material) and isolate well even after the 1170 K processing.

The top-electrode (either Re or Al) is then deposited {\it in situ} by use of UHV DC magnetron-sputtering after the
wafer is cooled to room-temperature ($T = 295$ K) in a 5 mPa oxygen background. The Al is deposited at a rate of 3 nm/min, and the argon sputter gas pressure is 0.7 Pa. For the Re top-electrode, we use xenon sputter gas instead of argon to avoid the creation of energetic neutral sputter-gas atoms, which act as an unintentional mill of the tunnel-barrier during the first few atomic layers of top-electrode deposition. This is because energetic neutrals are created when there is a large mismatch in atomic mass between the sputter gas and the target material \cite{Window}. The use of xenon instead of argon for Re sputtering reduces the fractional energy of neutrals from 0.42 to 0.03. According to RHEED and AFM (not shown), the Al and Re top-electrodes exhibit moderately textured in-plane crystalline order, but small $\sim 30$ nm grain-size due to the low 295 K deposition temperature. In the final step, figure \ref{fabcartoon}(d), the Re top-electrode is patterned by use of a 500 V SF$_6$ RIE at 2 Pa. If the top-electrode is Al, we use a 200 V argon ion mill at 0.4 Pa (oriented 20 degrees from substrate normal with sample rotation).

\section{Electrical Characterization} We measure the room-temperature ($T =$ 295 K) resistance of octagonal test junctions ranging in designed minimal diameter $d$ from 0.5 $\mu$m to 15 $\mu$m (area: 0.2 $\mu$m$^2$ to 186 $\mu$m$^2$). This provides three important pieces of information, including the process bias $d_0$. First, for medium and large junction sizes ($d >> d_0$), the RA-product should be flat when plotted versus designed area if there are no spurious transport channels at the perimeter of the junction. Second, by plotting RA-product versus electrical area $0.827(d-d_0)^2$ and adjusting $d_0$ so that we obtain a flat RA-product curve for small junction sizes ($d \sim d_0$), we extract $d_0$. This gives us information concerning how the actual size of the junction differs from the designed size. Third, if the superconducting gaps of the top and base-electrodes are known, the critical current density for the junctions in the superconducting regime can be calculated \cite{Ambegaokar}. This gives us feedback to adjust the RA-product by changing the tunnel-barrier deposition time for subsequent wafers.

\begin{figure}[ht] \centerline{\includegraphics[width = 6.98 cm]{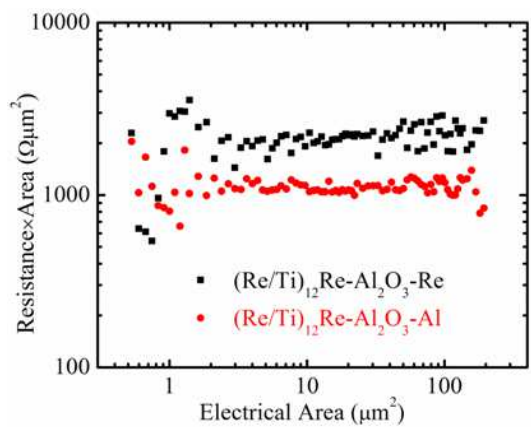}} \caption{Room-temperature ($T =$ 295 K)
resistance measurements of tunnel junctions with Re (black $\fullsquare$ curve) and Al (red \textcolor{red}{$\fullcircle$} curve) top-electrodes. Resistance measured at 100 nA bias current.} \label{RA} \end{figure}

Figure \ref{RA} shows a plot of RA-product versus electrical area for junctions with Re and Al top-electrodes. For both types, the curve is flat for medium and large junctions, so we expect no significant perimeter transport. Both types of junctions have a process bias of -0.3 $\mu$m, meaning that the junctions are 0.3 $\mu$m larger in diameter than designed. This agrees well with the SEM image in figure \ref{fabcartoon}(d), where a junction that was designed as 0.5 $\mu$m was measured to be 0.8 $\mu$m (area = 0.5 $\mu$m$^2$). In order to account for the observed spread in RA-product, we designed qubit circuits with various sized junctions, as described in \cite{Kline}. Based on measurements at $T = 295$ K, both types of top-electrodes appear favorable for use as Josephson junctions.

\begin{figure}[ht] \centerline{\includegraphics[width = 6.98 cm]{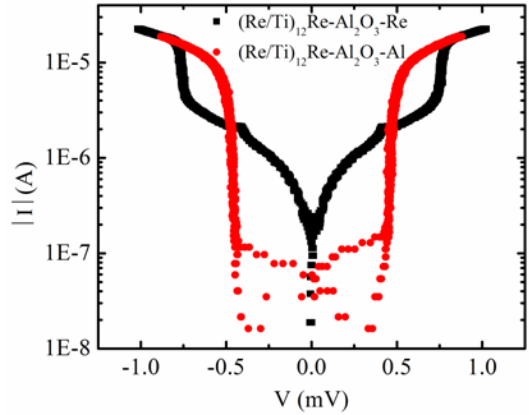}} \caption{Low-temperature ($T$ $\sim$ 50 mK) I-V curves for the Josephson junctions with Re (black $\fullsquare$ curve) and Al (red \textcolor{red}{$\fullcircle$} curve) top-electrodes. The superconducting branch is intentionally suppressed by a magnetic field oriented in the plane of the junction. We measure $\Delta_1+\Delta_2$ = $0.75$ meV for the junction with Re top-electrode, and $\Delta_1+\Delta_2 = 0.45$ meV for the junction with Al top-electrode.} \label{LTIV} \end{figure}

Low-temperature ($T$ $\sim$ 50 mK) measurements were then conducted for both the Re and Al top-electrode devices in an adiabatic demagnetization refrigerator using a commercial data-acquisition card and preamplifier. Figure \ref{LTIV} shows IV curves for two junctions of nominally equal area ($\sim$60 $\mu$m$^2$) and RA product ($\sim$2000 $\Omega$$\mu$m$^2$). While the normal-state resistances, i.e., the inverse slope of the curves above the superconducting gaps, are nearly the same, a dramatic difference in the subgap structure is observed. For the Re top-electrode junction, we see low subgap resistance $R_{sg} = 226$ $\Omega$, indicating some transport mechanism other than pure tunnelling. The subgap resistance is only five times higher than the normal-state resistance. The Al top-electrode junction shows a sharp corner, high subgap resistance and a re-trapping current that is limited by system noise, indicative of a high-quality junction \cite{Kirtley_subgap_PRL88}. We measured tunnel junctions ranging in size from $0.5-186$ $\mu$m$^2$ from five wafers with Re top-electrodes and six wafers with Al top-electrodes: all measurements exhibit the same qualitative behavior where the Re top-electrode junctions have low subgap resistance and the Al top-electrode junctions have high subgap resistance. We conclude that junctions made using Re top-electrodes have inherently poor subgap properties.

We measured the superconducting critical temperatures of the electrodes: $1.1$ K (Al), $2.5$ K (Re) and $2.4$ K ((Re/Ti)$_{12}$Re multilayer), corresponding to superconducting gaps $\Delta$ of $0.17$ meV (Al), $0.38$ meV (Re) and $0.36$ meV ((Re/Ti)$_{12}$Re), using BCS theory \cite{BCS}. The measured values $\Delta_1 + \Delta_2$ of $0.75$ meV for (Re/Ti)$_{12}$Re-Al$_2$O$_3$-Re and $0.45$ meV for (Re/Ti)$_{12}$Re-Al$_2$O$_3$-Al from the IV curves in figure \ref{LTIV} are in good agreement with theory.

\begin{figure}[ht] \centerline{\includegraphics[width = 6.98 cm]{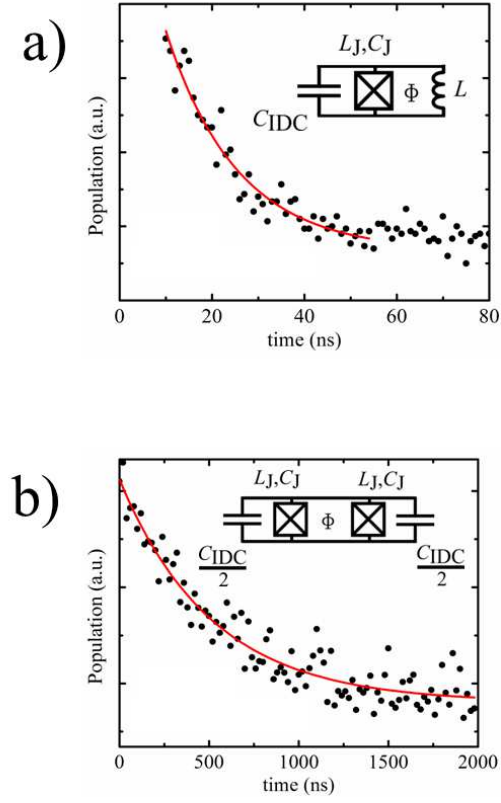}} \caption{Low-temperature ($T$ $<$ 100 mK) energy-relaxation measurements for (a) phase qubit with Re top-electrode: $T_1 = 15$ ns and, (b) transmon qubit with Al top-electrode: $T_1 = 500$ ns. Solid red line is exponential fit to data with decay time $T_1$. Insets show qubit circuit schematics where the Josephson junction inductance is $L_J$, the junction self capacitance is $C_J$, and the interdigitated capacitor has capacitance $C_{IDC}$. The phase qubit is shunted by loop inductance $L$. In each case, the qubit loop is threaded by on-chip adjustable magnetic flux $\Phi$.} \label{T1} \end{figure}

We also measured superconducting qubits made using the via process with both Re and Al top-electrodes. We first describe a flux-biased phase qubit with Re top-electrode. The circuit design is similar to \cite{Kline} with qubit state measurement performed using a DC SQUID. For a device with a 4 $\mu$m$^2$ qubit junction with capacitance $\sim200$ fF, critical current $= 2$ $\mu$A, 700 fF shunt (Re/Ti)$_{12}$Re interdigitated capacitor, loop inductance $L = 720$ pH, and 1 fF SiO$_x$ cross-over insulator, we measured an energy relaxation time $T_1$ = 15 ns, as shown in figure \ref{T1}(a). We hypothesize that $T_1$ is limited by the relatively low subgap resistance of the qubit junction with Re top-electrode; the classical $RC$ decay time for the qubit is $\tau=C R_{sg}\sim 2$ ns, where $C = 900$ fF is the total qubit capacitance. We measured two phase qubits from two wafers and both yielded similar results. We were unable to detect TLS splittings in the spectroscopy data due to the broad linewidth caused by the short relaxation time of these qubits.

A qubit fabricated with an Al top-electrode showed a much longer $T_1$ time of 500 ns, as shown in figure \ref{T1}(b). These data were taken from a transmission-line shunted plasma oscillation qubit (transmon) with dispersive qubit state readout using a half-wave resonator \cite{Koch}. The total qubit capacitance is given by two $1$ $\mu$m$^2$ junctions with $\sim100$ fF capacitance (critical current = 0.1 $\mu$A), $60$ fF shunt (Re/Ti)$_{12}$Re interdigitated capacitor, and 1 fF cross-over SiN$_x$ insulator. The half-wave resonator frequency is 8.3 GHz and the $T_1$ measurement was performed at the 7.3 GHz flux ``sweet spot''. We measured two transmon qubits from one wafer and both yielded similar results. Based on the qubit-resonator coupling of $ g/2\pi = 85$ MHz, qubit-resonator detuning $ \Delta/2\pi = 1$ GHz, and resonator photon loss rate $\kappa/2\pi = 0.8$ MHz: the Purcell effect limit on $T_1$ is $(\Delta/g^2)/\kappa = 27$ $\mu$s, so our devices are not limited by the Purcell effect. We observed three TLS splittings in the spectroscopy measurement over a 0.5 GHz range (not shown), with maximum splitting size = 7 MHz.

Table \ref{Participation} shows an analysis of the loss in each element of the transmon circuit: Josephson junction, interdigitated capacitor, and SiN$_x$ insulator. The participation ratio of each element is given by $p_i = C_i/C_{tot}$, where $C_i$ is the capacitance of element $i$ and $C_{tot}$ is the total qubit capacitance. The contributed loss is given by the microwave dielectric loss tangent (tan $\delta$) times $P_i$. Here we consider only the low-power loss tangent, i.e., the loss tangent measured when the TLS are unsaturated by the applied electric field \cite{Martinis:loss}. We use independently measured values of tan $\delta$ for the interdigitated capacitor and the SiN$_x$ insulator. We use the measured $T_1 = 500$ ns to calculate the total loss tangent of the transmon as $4.3 \times10^{-5}$ through $T_1 = 1/(2 \pi f_r$tan $\delta)$, where $f_r$ is the 7.3 GHz resonance frequency. We find that the performance of the qubit is limited primarily by loss in the Josephson junction and the interdigitated capacitor. Other loss mechanisms, such as non-equilibrium quasiparticles, are not considered in this analysis.

\begin{table}[ht]
\caption{Transmon loss analysis. The capacitance of element $i$ is $C_i$, the participation ratio of element $i$ is $P_i$, the loss tangent is tan $\delta$, and the contributed loss is given by tan $\delta \times P_i$.}
\begin{indented}
\lineup
\item[]\begin{tabular}{cccccc}
\br
Element&         $C_i$ (fF)&         $P_i$ (\%)&         tan $\delta$&        Contributed loss \\
\mr
Junction&               100&                 62.1&              $3.5\times10^{-5}$&         $2.2\times10^{-5}$ \\
IDC&              $\0$60&                  37.3&              $4.0\times10^{-5}$&         $1.5\times10^{-5}$ \\
SiN$_x$&            $\0$$\0$1&                   $\0$0.6&               $1.0\times10^{-3}$&         $6.2\times10^{-6}$ \\
\mr
       &              &                      &               Total loss&               $4.3 \times10^{-5}$ \\
\br
\label{Participation}
\end{tabular}
\end{indented}
\end{table}

\section{Conclusions}
We have presented a recipe for the fabrication of sub-micrometer epitaxial Josephson junctions with Al$_2$O$_3$ tunnel-barriers. The substrate crystallinity has been improved by a furnace anneal, and the base-electrode has been smoothed through the use of a (Re/Ti)$_{12}$Re multilayer base-electrode. The epitaxial Al$_2$O$_3$ tunnel-barrier is deposited at the bottom of a via in either SiO$_x$ or SiN$_x$. The top-electrodes are made from either Re or Al.

We find that Josephson junctions fabricated using the via process with Re top-electrodes have low subgap resistance and phase qubit energy relaxation time $T_1 = 15$ ns. This energy relaxation time is much smaller than the $T_1 = 500$ ns measured on a large area ($49$ $\mu$m$^2$) epitaxial Re-Al$_2$O$_3$-Al phase qubit \cite{Kline} fabricated using a trilayer process in the same laboratory as the devices studied in this work and also the best amorphous-barrier phase qubit with $T_1 = 600$ ns \cite{Martinis-QIP}. We find that our Al top-electrode devices have a high junction subgap resistance and transmon qubit energy relaxation time $T_1 = 500$ ns. We note that the best amorphous-barrier transmon, with junction area $\sim 0.1$ $\mu$m$^2$, has $T_1 = 2000$ ns for operation at $f \sim 6$ GHz \cite{Houck}.

\begin{ack} We gratefully acknowledge the fabrication assistance of Farnaz Farhoodi. The devices were fabricated at NIST. Room-temperature junction measurements were performed at MIT Lincoln Laboratory. The phase qubits were measured at The Hebrew University of Jerusalem and the transmons were measured at Raytheon BBN Technologies. This work was funded by the NIST Quantum Information initiative and the Office of the Director of National Intelligence (ODNI) through the Intelligence Advanced Research Projects Activity (IARPA).  All statements of fact, opinion, or conclusions contained herein are those of the authors and should not be construed as representing the official views or policies of ODNI or IARPA. \end{ack}

\section{References}
\bibliography{ViaProcess}

\begin{thebibliography}{10}

\bibitem{Clarke:review}
J.~Clarke and F.~K. Wilhelm.
\newblock {Superconducting quantum bits}.
\newblock {\em {Nature}}, {453}({7198}):{1031--1042}, Jun 19 {2008}.

\bibitem{Simmonds:TLS}
R.~W. Simmonds, K.~M. Lang, D.~A. Hite, S.~Nam, D.~P. Pappas, and J.~M.
  Martinis.
\newblock {Decoherence in Josephson Phase Qubits from Junction Resonators}.
\newblock {\em Phys. Rev. Lett.}, 93(7):077003, Aug 2004.

\bibitem{Martinis:loss}
J.~M. Martinis, K.~B. Cooper, R.~McDermott, M.~Steffen, M.~Ansmann, K.~D.
  Osborn, K.~Cicak, S.~Oh, D.~P. Pappas, R.~W. Simmonds, and Clare~C. Yu.
\newblock {Decoherence in Josephson qubits from dielectric loss}.
\newblock {\em Phys. Rev. Lett.}, 95(21):210503, Nov 2005.

\bibitem{Kline}
J.~S. Kline, H.~Wang, S.~Oh, J.~M. Martinis, and D.~P. Pappas.
\newblock Josephson phase qubit circuit for the evaluation of advanced tunnel
  barrier materials.
\newblock {\em Supercond. Sci. Tech.}, 22(1):015004, 2009.

\bibitem{Dolan77}
G.~J. Dolan.
\newblock {Offset masks for lift-off photoprocessing}.
\newblock {\em Appl. Phys. Lett.}, 31(5):337--339, 1977.

\bibitem{Kim:chgqubit}
Z.~Kim, V.~Zaretskey, Y.~Yoon, J.~F. Schneiderman, M.~D. Shaw, P.~M.
  Echternach, F.~C. Wellstood, and B.~S. Palmer.
\newblock {Anomalous avoided level crossings in a Cooper-pair box spectrum}.
\newblock {\em Phys. Rev. B}, 78(14):144506, Oct 2008.

\bibitem{SchreierPRB08}
J.~A. Schreier, A.~A. Houck, Jens Koch, D.~I. Schuster, B.~R. Johnson, J.~M.
  Chow, J.~M. Gambetta, J.~Majer, L.~Frunzio, M.~H. Devoret, S.~M. Girvin, and
  R.~J. Schoelkopf.
\newblock {Suppressing charge noise decoherence in superconducting charge
  qubits}.
\newblock {\em Phys. Rev. B}, 77(18):180502, May 2008.

\bibitem{Mooij}
J.~E. Mooij, T.~P. Orlando, L.~Levitov, L.~Tian, C.~H. van~der Wal, and
  S.~Lloyd.
\newblock Josephson persistent-current qubit.
\newblock {\em Science}, 285(5430):1036--1039, 1999.

\bibitem{Steffen:Zqubit}
M.~Steffen, S.~Kumar, D.~P. DiVincenzo, J.~R. Rozen, G.~A. Keefe, M.~B.
  Rothwell, and M.~B. Ketchen.
\newblock High-coherence hybrid superconducting qubit.
\newblock {\em Phys. Rev. Lett.}, 105(10):100502, Sep 2010.

\bibitem{Steffen:overlap}
M.~Steffen, M.~Ansmann, R.~McDermott, N.~Katz, Radoslaw~C. Bialczak, E.~Lucero,
  M.~Neeley, E.~M. Weig, A.~N. Cleland, and J.~M. Martinis.
\newblock State tomography of capacitively shunted phase qubits with high
  fidelity.
\newblock {\em Phys. Rev. Lett.}, 97(5):050502, Aug 2006.

\bibitem{Weides_Trilayer}
M.~Weides, R.~C. Bialczak, M.~Lenander, E.~Lucero, Matteo Mariantoni,
  M.~Neeley, A.~D. O'Connell, D.~Sank, H.~Wang, J.~Wenner, T.~Yamamoto, Y.~Yin,
  A.~N. Cleland, and J.~Martinis.
\newblock {Phase qubits fabricated with trilayer junctions}.
\newblock {\em Supercond. Sci. and Tech.}, 24(5):055005, May 2011.

\bibitem{Oh:junctions}
S.~Oh, K.~Cicak, R.~McDermott, K.~B. Cooper, K.~D. Osborn, R.~W. Simmonds,
  M.~Steffen, J.~M. Martinis, and D.~P. Pappas.
\newblock {Low-leakage superconducting tunnel junctions with a single-crystal
  Al2O3 barrier}.
\newblock {\em Supercond. Sci. Technol.}, 18(10):1396, 2005.

\bibitem{Oh:TLS}
S.~Oh, K.~Cicak, J.~S. Kline, M.~A. Sillanp\"a\"a, K.~D. Osborn, J.~D.
  Whittaker, R.~W. Simmonds, and D.~P. Pappas.
\newblock Elimination of two level fluctuators in superconducting quantum bits
  by an epitaxial tunnel barrier.
\newblock {\em Phys. Rev. B}, 74(10):100502, Sep 2006.

\bibitem{Berggren}
K.~K. Berggren, E.~M. Macedo, D.~A. Feld, and J.~P. Sage.
\newblock {Low Tc superconductive circuits fabricated on 150-mm-diameter wafers
  using a doubly planarized Nb/AlOx/Nb process}.
\newblock {\em IEEE Transactions on Applied Superconductivity}, 9(2):3271 --
  3274, Jun 1999.

\bibitem{Bertness}
K.~A. Bertness, A.~W. Sanders, D.~M. Rourke, T.~E. Harvey, A.~Roshko, J.~B.
  Schlager, and N.~A. Sanford.
\newblock {Controlled nucleation of GaN nanowires grown with molecular beam
  epitaxy}.
\newblock {\em Advanced Functional Materials}, 20(17):2911--2915, 2010.

\bibitem{Cuccureddu}
F.~Cuccureddu, S.~Murphy, I.~V. Shvets, M.~Porcu, H.~W. Zandbergen, N.~S.
  Sidorov, and S.~I. Bozhko.
\newblock {Surface morphology of c-plane sapphire (alpha-alumina) produced by
  high temperature anneal}.
\newblock {\em Surface Science}, {604}({15-16}):{1294--1299}, {Aug 15} {2010}.

\bibitem{Yoshimoto}
M.~Yoshimoto, T.~Maeda, T.~Ohnishi, H.~Koinuma, O.~Ishiyama, M.~Shinohara,
  M.~Kubo, R.~Miura, and A.~Miyamoto.
\newblock Atomic-scale formation of ultrasmooth surfaces on sapphire substrates
  for high-quality thin-film fabrication.
\newblock {\em Appl. Phys. Lett.}, 67(18):2615, 1995.

\bibitem{Oh:rhenium}
S.~Oh, D.~A. Hite, K.~Cicak, K.~D. Osborn, R.~W. Simmonds, R.~McDermott, K.~B.
  Cooper, M.~Steffen, J.~M. Martinis, and D.~P. Pappas.
\newblock Epitaxial growth of rhenium with sputtering.
\newblock {\em Thin Solid Films}, 496(2):389, 2006.

\bibitem{Du}
J.~Du, A.~D.~M. Charles, K.~D. Petersson, and E.~W. Preston.
\newblock {Influence of Nb film surface morphology on the sub-gap leakage
  characteristics of Nb/AlOx-Al/Nb Josephson junctions}.
\newblock {\em Supercond. Sci. Technol.}, {20}({11, Sp. Iss. SI}):{S350--S355},
  Nov {2007}.

\bibitem{Welander}
P.~B. Welander, T.~J. McArdle, and J.~N. Eckstein.
\newblock {Reduced leakage current in Josephson tunnel junctions with
  codeposited barriers}.
\newblock {\em Appl. Phys. Lett.}, 97(23):233510, 2010.

\bibitem{Kohlstedt}
H.~Kohlstedt, F.~Konig, P.~Henne, N.~Thyssen, and P.~Caputo.
\newblock {The role of surface roughness in the fabrication of stacked
  Nb/Al-AlOx/Nb tunnel junctions}.
\newblock {\em J. Appl. Phys.}, 80(9):5512, 1996.

\bibitem{Kominami}
S.~Kominami, H.~Yamada, N.~Miyamoto, and K.~Takagi.
\newblock {Effects of underlayer roughness on Nb/AlOx/Nb junction
  characteristics}.
\newblock {\em IEEE Transactions on Applied Superconductivity}, 3(1):2182
  --2186, Mar 1993.

\bibitem{Barner1}
J.~Barner and S.~Ruggiero.
\newblock {RF sputter-deposited aluminum-oxide films as high quality artificial
  tunnel barriers}.
\newblock {\em IEEE Transactions on Magnetics}, 23(2):854 -- 858, Mar 1987.

\bibitem{Barner:tunneling}
J.~B. Barner and S.~T. Ruggiero.
\newblock {Tunneling in artificial Al2O3 tunnel barriers and Al2O3-metal
  multilayers}.
\newblock {\em Phys. Rev. B}, 39(4):2060--2071, Feb 1989.

\bibitem{Window}
B.~Window.
\newblock Removing the energetic neutral problem in sputtering.
\newblock {\em 39th National Symposium of the American Vacuum Society},
  11(4):1522--1527, 1993.

\bibitem{Ambegaokar}
V.~Ambegaokar and A.~Baratoff.
\newblock Tunneling between superconductors.
\newblock {\em Phys. Rev. Lett.}, 10(11):486--489, Jun 1963.

\bibitem{Kirtley_subgap_PRL88}
J.~R. Kirtley, C.~D. Tesche, W.~J. Gallagher, A.~W. Kleinsasser, R.~L.
  Sandstrom, S.~I. Raider, and M.~P.~A. Fisher.
\newblock {Measurement of the intrinsic subgap dissipation in Josephson
  junctions}.
\newblock {\em Phys. Rev. Lett.}, 61(20):2372--2375, November 1988.

\bibitem{BCS}
J.~Bardeen, L.~N. Cooper, and J.~R. Schrieffer.
\newblock Theory of superconductivity.
\newblock {\em Phys. Rev.}, 108(5):1175--1204, Dec 1957.

\bibitem{Koch}
J.~Koch, T.~M. Yu, J.~Gambetta, A.~A. Houck, D.~I. Schuster, J.~Majer,
  A.~Blais, M.~H. Devoret, S.~M. Girvin, and R.~J. Schoelkopf.
\newblock {Charge-insensitive qubit design derived from the Cooper pair box}.
\newblock {\em Phys. Rev. A}, 76(4):042319, Oct 2007.

\bibitem{Martinis-QIP}
John Martinis.
\newblock Superconducting phase qubits.
\newblock {\em Quantum Information Processing}, 8:81--103, 2009.
\newblock 10.1007/s11128-009-0105-1.

\bibitem{Houck}
A.~A. Houck, J.~A. Schreier, B.~R. Johnson, J.~M. Chow, Jens Koch, J.~M.
  Gambetta, D.~I. Schuster, L.~Frunzio, M.~H. Devoret, S.~M. Girvin, and R.~J.
  Schoelkopf.
\newblock Controlling the spontaneous emission of a superconducting transmon
  qubit.
\newblock {\em Phys. Rev. Lett.}, 101:080502, Aug 2008.

\end{thebibliography}

\newpage

\end{document}